\documentclass[12pt]{article}   %% LaTeX 2e (preferred)
\usepackage{opex3} %% do not use with REVTeX4
\usepackage[draft]{hyperref} %% optional
\usepackage{amsmath}
\usepackage{epsfig}

\newcommand{\SiN}{\text{Si}_\text{3}\text{N}_\text{4}}
\newcommand{\mum}{~\mu{\text{m}}}

\newcommand{\gom}{g_\text{OM}}
\newcommand{\lom}{L_\text{OM}}

\begin{document}

%\twocolumn[

\title{Slot-mode-coupled optomechanical crystals}

%% For REVTeX it is possible to automate superscript and e-mail callouts with the superscriptaddress option; see REVTeX4 documentation.

\author{Marcelo Davan\c co$^{* 1,2}$, Jasper Chan$^2$, Amir H. Safavi-Naeini$^2$, Oskar Painter$^2$, and Kartik Srinivasan$^1$}
\address{$^1$Center for Nanoscale Science and Technology,
National Institute of Standards and Technology, Gaithersburg, MD\\
$^2$Thomas J. Watson, Sr., Laboratory of Applied Physics, California
Institute of Technology, Pasadena, CA 91125\\$^*$Corresponding
author: mdavanco@nist.gov}

\begin{abstract} We present a design methodology and analysis of
a cavity optomechanical system in which a localized GHz frequency
mechanical mode of a nanobeam resonator is evanescently coupled to a
high quality factor ($Q>10^6$) optical mode of a separate nanobeam
optical cavity.  Using separate nanobeams provides flexibility,
enabling the independent design and optimization of the optics and
mechanics of the system. In addition, the small gap ($\approx$25~nm)
between the two resonators gives rise to a slot mode effect that
enables a large zero-point optomechanical coupling strength to be
achieved, with $g/2\pi>300$~kHz in a Si$_3$N$_4$ system at 980~nm
and $g/2\pi\approx900$~kHz in a Si system at 1550~nm. The fact that
large coupling strengths to GHz mechanical oscillators can be
achieved in $\SiN$ is important, as this material has a broad
optical transparency window, which allows operation throughout the
visible and near-infrared. As an application of this platform, we
consider wide-band optical frequency conversion between 1300~nm and
980~nm, using two optical nanobeam cavities coupled on either side
to the breathing mode of a mechanical nanobeam resonator.
\end{abstract}


%]

%\maketitle %% null function with osajnl.sty

%\linenumbers

%\bibliography{KS_bib_2012_5_31}
\bibliographystyle{osa}

\section{Introduction}

Recent experiments using radiation pressure effects in cavity
optomechanical systems~\cite{ref:Kippenberg_Vahala_OE}, such as
electromagnetically-induced
transparency~\cite{ref:Weis_Kippenberg,ref:safavi-naeini4} and
ground state
cooling~\cite{ref:Teufel_ground_state,ref:Chan_Painter_ground_state},
as well as theoretical proposals for using such optomechanical
systems in quantum information processing
applications~\cite{ref:safavi-naeini3,ref:Tian_Wang_PRA,ref:Chang_Painter,ref:Stannigel_pra,ref:Ludwig},
generally rely upon the achievement of large zero-point
optomechanical coupling rates and low optical and mechanical
dissipation rates.  They also typically require operation in the
resolved sideband limit in which the mechanical resonance frequency
must exceed the optical dissipation rate. One-dimensional
optomechanical
crystals~\cite{ref:eichenfield2,ref:safavi-naeini4,ref:Chang_Painter},
in which a periodic patterning is applied to a doubly clamped
nanobeam in order to simultaneously localize GHz phonons and
near-infrared (200~THz) photons, are one promising architecture for
future work in quantum cavity optomechanics.

While the use of a single nanobeam in such work has many advantages,
separation of the optical cavity and mechanical resonator can afford
greater levels of flexibility. This can be useful in applications in
which multiple optical modes are coupled to the same mechanical
resonance, for example, to achieve wavelength
conversion~\cite{ref:safavi-naeini3,ref:Tian_Wang_PRA,ref:hill_WLC,ref:dong_wlc}
or in converting traveling wave phonons to traveling wave
photons~\cite{ref:safavi-naeini3}.  One disadvantage of separating
the mechanical resonator and optical cavity is that the
optomechanical coupling rate generally tends to be smaller, given
the reduced overlap between the mechanical and optical modes. To
overcome this, one can focus on optical modes formed within the gap
between the optical and mechanical resonator. Such air-slot type
modes~\cite{Almeida.OL.04}, which display a high electric field
concentration in the air gap, have been used to achieve large
optomechanical
couplings~\cite{ref:eichenfield1,ref:Roh,ref:Lin5,ref:Wiederhecker_Lipson},
and sub-wavelength effective optomechanical
lengths~\cite{ref:safavi-naeini2}.  In these demonstrations,
however, the mechanical resonators and optical cavities were not
separated, and mechanical resonance frequencies were typically
$<150$~MHz.  For operation in the resolved sideband regime, higher
frequencies are desirable, given that high-$Q$ cavities in materials
like Si and $\SiN$ typically have optical dissipation rates that are
$>100$~MHz.

\begin{figure}[]
\centerline{\includegraphics[width=8.5cm,trim=10mm 0mm 40mm
0mm]{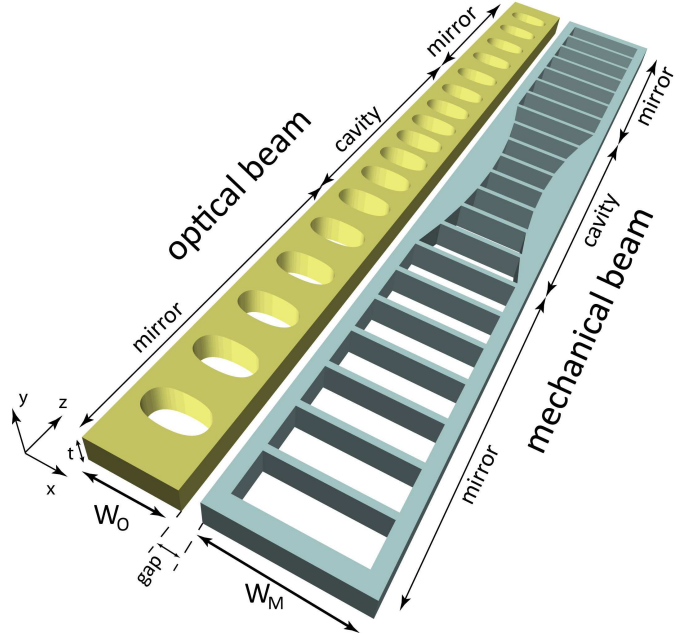}} \caption{Schematic of double beam optomechanical
resonator. } \label{FIG_F}
\end{figure}

Here, we present the design of a cavity optomechanical system in
which a GHz frequency localized mode in a mechanical nanobeam
resonator is coupled to a localized 300 THz optical mode in a
separate optical nanobeam resonator (Fig.~\ref{FIG_F}). We show that
the slot mode effect can increase optomechanical coupling rates with
respect to those found in a single nanobeam optomechanical system,
and consider applications of this approach to instances in which the
separated mechanical and optical resonators are particularly
advantageous, such as wide-band optical wavelength
conversion~\cite{ref:safavi-naeini3}. We consider both $\SiN$ and Si
systems, although our primary focus is on $\SiN$. In particular, the
$\SiN$ system has many potential advantages for applications in
quantum cavity optomechanics: optical $Q>10^6$ has been demonstrated
over a wide range of wavelengths in the visible and near-infrared in
$\SiN$~\cite{ref:Barclay8,ref:Gondarenko,ref:Hosseni_Adibi}. This
can allow compatibility with quantum systems based on trapped atoms,
ions, and semiconductor quantum dots, whose relevant optical
transitions are often below 1000~nm (where Si is opaque).
Furthermore, these stoichiometric Si$_3$N$_4$ films under tensile
stress have been shown to support high mechanical quality factors
($\approx10^6$), both in commercial thin
films~\cite{ref:Zwickl_Harris} and in nanofabricated
devices~\cite{ref:Verbridge,ref:Fong_Tang,ref:Krause_Painter}.
Finally, as Si$_3$N$_4$ does not exhibit the two-photon absorption
and subsequent free-carrier absorption and dispersion that silicon
does~\cite{ref:Barclay7}, higher optical powers are usable in this
system, potentially increasing the range of pump-enhanced
optomechanical coupling values that can be achieved. Our geometry is
predicted to support GHz mechanical frequencies, optical $Q>10^6$,
and a zero-point optomechanical coupling strength $g/2\pi>300$~kHz.
Such values should in principle enable strong radiation pressure
driven phenomena.

The proposed geometry, shown in Fig.~\ref{FIG_F}, consists of two
suspended parallel dielectric beams of refractive index $n_d$,
widths $W_{O}$ and $W_{M}$ and thickness $t$, separated by a gap of
width $w_{gap}$. Both beams are etched through to form
quasi-periodic 1D structures in the $z$-direction. The "optical"
beam's 1D lattice is modulated so that a quasi-TE ($E_y=0$ across
the center of the $\SiN$ layer at $y = 0$) confined optical mode is
supported at an optical frequency $f_o$. Optical confinement along
the beam is provided by a 1D photonic bandgap, while total internal
reflection provides confinement in the $x$ and $y$
directions~\cite{ref:eichenfield2,ref:Quan1,ref:Quan2}. With a
sufficiently small spacing $w_{gap}$, the quasi-TE mode consists of
an air-gap resonance as in~\cite{Almeida.OL.04}, with a strongly
concentrated, almost completely $x$-oriented electric field between
the two beams. The "mechanical" beam is designed to support a
mechanical resonance of frequency $f_m$, confined in $z$, that
displaces the boundary $B$ facing the optical resonator, thereby
closing the gap and thus shifting the optical resonance frequency
$f_o$. It is important to stress that the optical cavity is actually
formed by the "optical" beam and the "mechanical" beam's closest arm
(indicated in Fig.~\ref{FIG_F}), so that optical resonances depend
strongly on the gap width. Nevertheless, optical and mechanical
resonances can be tuned with great independence, as we show in the
following sections.  Sections~\ref{section:background} and
\ref{section:single_nanobeam} present background information on the
simulations performed and consider the case of a single Si$_3$N$_4$
nanobeam optomechanical cavity, as a baseline reference.
Section~\ref{section:gap_nanobeams} describes the design of the
slot-mode-coupled optical and mechanical nanobeam resonators,
emphasizing the design approach for achieving high-$Q$ optical
modes, GHz mechanical modes, and large optomechanical coupling.
Finally, section~\ref{section:Applications} considers a few
applications of this platform.

\section{Background}
\label{section:background}
The shift in the frequency $f_o$ of a
particular optical resonance due to displacement of the
nanostructure boundaries produced by a mechanical resonance at
frequency $f_m$ is quantified by the optomechanical coupling $\gom =
\partial \omega_o/\partial x=\omega_o/\lom$; here, $x$ is the cavity
boundary displacement, $\omega_o=2\pi
f_o$ and $\lom$ is an effective optomechanical interaction
length~\cite{ref:eichenfield3}. The effective length $\lom$ can be
estimated via the perturbative expression~\cite{ref:Johnson6}
\begin{equation}
\label{eq:Lom} \lom = \frac{2\int {dV \epsilon\left|\mathbf{E}
\right|^2}}{\int{dA\left( \left| \mathbf{Q}\cdot \mathbf{n}\right|
\right)\left( \Delta \epsilon\left| \mathbf{E}_\parallel
\right|^2-\Delta(\epsilon^{-1}) \left| \mathbf{D}_\perp \right|
^2\right)} }.
\end{equation}
Here, $\mathbf{E}$ and $\mathbf{D}$ are the modal electric and
electric displacement fields, respectively, $\Delta \epsilon =
\epsilon_{diel.}-\epsilon_{air}$, $\Delta(\epsilon^{-1}) =
\epsilon_{diel.}^{-1}-\epsilon_{air}^{-1}$, and $\epsilon_{diel.,
air}$ are the permittivities of the nanobeam material and air. The
mass displacement due to the mechanical resonance is given by
$\mathbf{Q}$, and the normal surface displacement at the structure
boundaries is $\left| \mathbf{Q}\cdot \mathbf{n}\right|$, where
$\mathbf{n}$ is the surface normal. The integral in the denominator
is performed over the entire surface of the nanostructure.

The optomechanical coupling $\gom$ can be converted into a pure
coupling rate $g$ between the optical and mechanical resonances:
$g=x_{zpf}\cdot \gom$, where $x_{zpf}=\sqrt{\hbar/2m\omega_m}$ is
the zero point fluctuation amplitude for mechanical
displacement~\cite{ref:safavi-naeini3} and $m$ is the motional mass
of the mechanical resonance at frequency $\omega_m$. The motional
mass can be obtained from the displacement $\mathbf{Q}$ and the
nanobeam material density $\rho$ by
$m=\rho\int{dV\left(\frac{|\mathbf{Q}|}{\text{max}(|\mathbf{Q}|)}\right)^2}$.

We point out that eq.(\ref{eq:Lom}) only quantifies frequency
changes due to boundary distortions. An additional, potentially
important contribution to the optomechanical coupling comes from the
photo-elastic effect, which corresponds to local, stress-induced
changes of the refractive index. This contribution can be quite
significant, and was in fact recently been exploited to demonstrate,
in combination with the moving-boundary contribution, optomechanical
coupling rates in excess of $1$~MHz in silicon nanobeam
optomechanical crystals~\cite{ref:chan_arxiv_2012}. Giant
enhancement of Brillouin scattering in suspended silicon waveguides,
resulting from both photo-elastic and radiation pressure
optomechanical coupling, has also been
predicted~\cite{ref:rakich_PRX}. In the present work, only
moving-boundary contributions were taken into account.

Optical and mechanical resonator modes are obtained using the finite
element method, as described in ~\cite{ref:eichenfield1}, and are
used in the expressions above to yield the optomechanical coupling
rates.

\section{$\SiN$ nanobeam}
\label{section:single_nanobeam} 
\begin{figure}[]
\centerline{\includegraphics[width=13.5cm,trim=45mm 60mm 45mm
0mm]{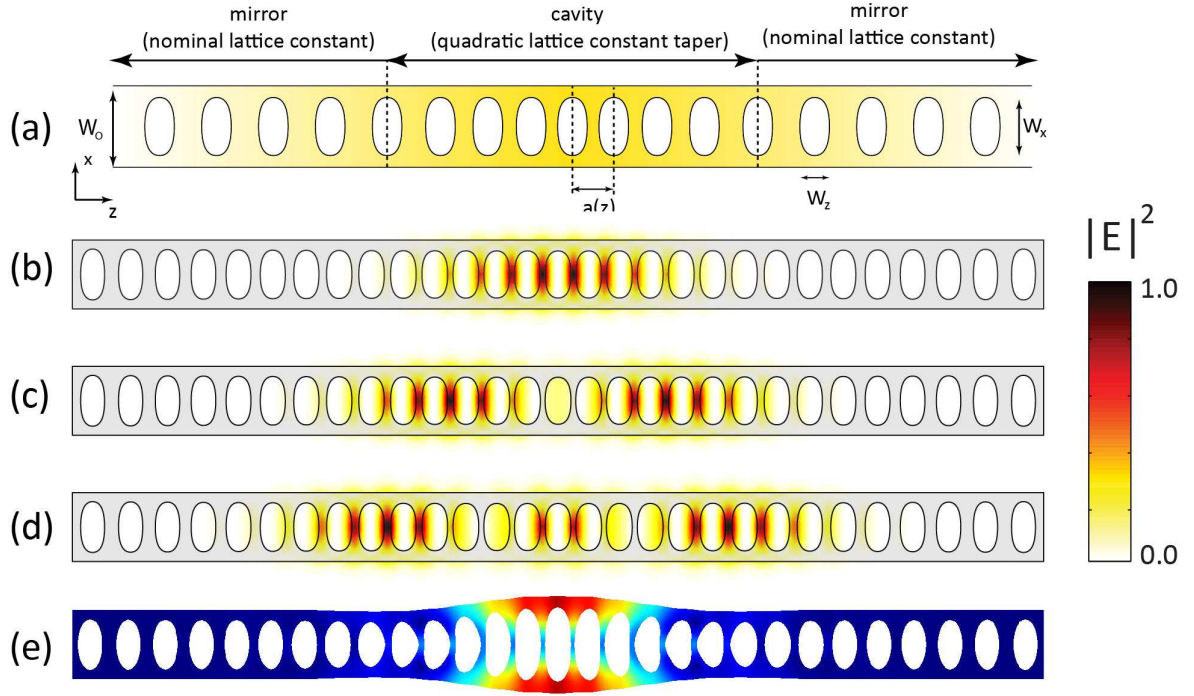}}\caption{(a) Nanobeam optomechanical
crystal. (b)-(d): first, second and third order optical resonances.
(e) Mechanical resonance. } \label{FIG:single_nanobeam}
\end{figure}

To gain some initial insight into
the optomechanical rates achievable with a low index contrast
nanobeam with 'standard' (i.e., single nanobeam) geometry, we start
by investigating an optical cavity composed of a 1D array of air
holes etched on a $\SiN$ nanobeam, as show in
Fig.~\ref{FIG:single_nanobeam}(a)~\cite{ref:Khan1}. We seek to
achieve the largest possible optomechanical coupling rate $g$ and
optomechanical coupling,  $\gom$. Our cavity was designed to support
the optical mode at a wavelength $\lambda=960$~nm shown in
Fig.~\ref{FIG:single_nanobeam}(b). The 980~nm wavelength band is
technologically important due to the availability of triggered
single photon sources based on InAs quantum
dots~\cite{ref:Shields_NPhot} and silicon based single photon
detectors which operate within this
range~\cite{ref:Hadfield_nphoton_09}.  Together with the optical
mode, this cavity supports confined mechanical modes co-located with
the optical resonance. The fundamental breathing mechanical mode
(FBM) at $f=3.2$~GHz, shown in Fig.~\ref{FIG:single_nanobeam}(e),
with motional mass $m=570$~fg, displays the highest optomechanical
coupling rate $g$ to the optical resonance in
Fig.~\ref{FIG:single_nanobeam}(b).

To reach this design, the procedure outlined in
Section~\ref{section:optical} was initially employed to obtain a
high $Q$ optical mode. The resulting cavity displayed a quadratic
variation of the lattice constant (the spacing between holes) from
the center to the edges~\cite{ref:Quan1}, and supported, together
with the optical mode, a localized breathing mechanical mode. The
cavity geometry was then optimized via a nonlinear minimization code
to yield a maximized optomechanical coupling rate $g/2\pi=133.6$~kHz
($\lom=5.1~\mum$), while maintaining a high optical quality factor
$Q>1\times10^6$. The optimization was realized by allowing the
aspect ratio of the elliptical holes (vertical axis over horizontal
axis) to vary quadratically from the cavity center, then searching a
2D space of aspect ratios at the center and at the edge of the
cavity. Reducing the aspect ratio at the cavity center proved to be
beneficial in reducing the effective length $L_{OM}$, by reducing
the mechanical mode volume and promoting better overlap between the
optical and mechanical modes. The $Q$ factor was not strongly
affected, most likely because the hole areas were maintained.

The nanobeam also supports two additional high order modes, shown in
Figs.~\ref{FIG:single_nanobeam}(c) and (d), at wavelengths $999$~nm
and $1036$~nm, both with $Q>1\times10^6$, which couple to the
mechanical mode in Fig.~\ref{FIG:single_nanobeam}(e) with
$g/2\pi=55.2$~kHz and $g/2\pi=17.5$~kHz, respectively.

\subsection*{Wavelength conversion}
\label{section:wlc_single} A cavity optomechanical system supporting
two optical resonances at different wavelengths and a shared
mechanical resonance may be used for performing wavelength
conversion of classical or quantum optical
signals~\cite{ref:safavi-naeini3,ref:Tian_Wang_PRA,ref:hill_WLC,ref:dong_wlc}.
In this photon-phonon translator scheme, a signal tuned to one of
the optical resonances is transduced to the shared mechanical
resonance, and from the latter to a second optical resonance at a
different wavelength. These conditions are achieved in the single
nanobeam design above, where the three displayed optical resonances
couple to the breathing mechanical mode. The wavelength separation
between input and converted signals is $\approx76$~nm, given by the
separation between the fundamental and third order optical modes.
Wider separations can in principle be achieved in larger cavities,
however, given the requirement of small optical and mechanical modal
volumes for increased optomechanical coupling, $g$ is expected to
drop~\cite{ref:Safavi-Naeini1}. Ultimately, the achievable spectral
separation in single nanobeam designs has an upper bound given by
the achievable photonic bandgap width, which increases with the
refractive index contrast. Nanobeams based on $\SiN$, with
refractive indices $n\approx2.0$, are considerably more limited than
Si nanobeams ($n\approx3.5)$.

\section{Slot mode nanobeams}
\label{section:gap_nanobeams}
\subsection{Optical cavity design}
\label{section:optical} 
\begin{figure}[]
\centerline{\includegraphics[width=13.5cm]{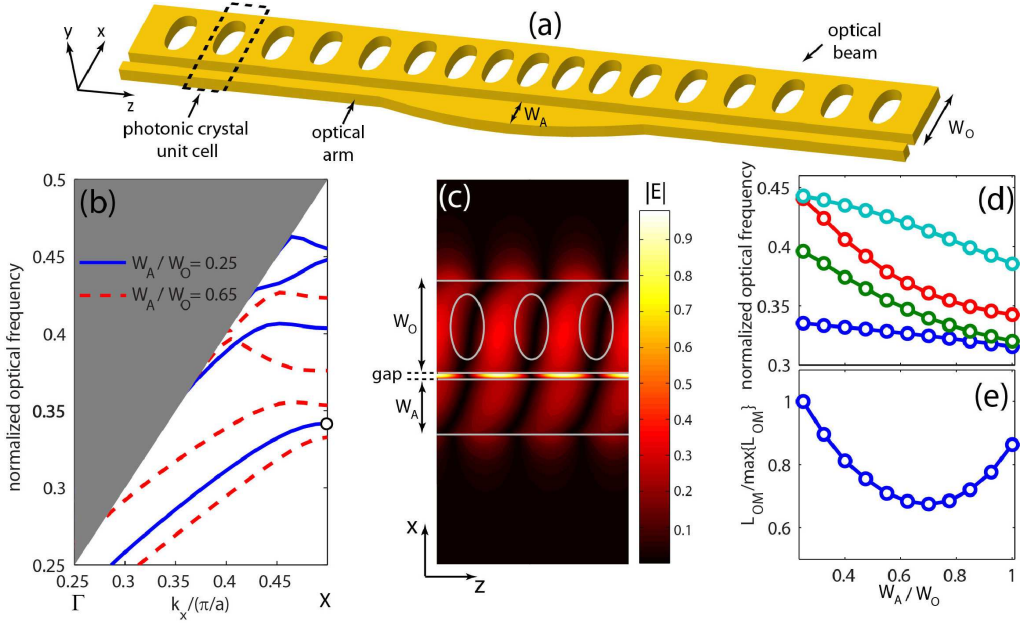}}\caption{(a)
Simplified nanobeam geometry with only the optical arm of the
mechanical resonator. (b) Photonic bands for TE ($E_z=0$ at the
$z=0$ plane) mode of the optomechanical crystal unit cell indicated
in (a), for $W_A/W_O=0.25$ and $W_A/W_O=0.65$. (c) Normalized
electric field amplitude for the fundamental TE mode band at the $X$
point (open circle in (b)). (d) Normalized frequency bands for the
first four TE photonic Bloch modes at the $X$-point, and (e)
normalized effective length for the fundamental TE mode, as
functions of $W_A/W_O$.} \label{FIG_B}
\end{figure}

We start the design of a $\SiN$ optical
cavity for operation in the 980~nm band by considering the geometry
in Fig.~\ref{FIG_B}(a). This is a version of the geometry in
Fig.~\ref{FIG_F} in which the mechanical resonator is stripped of
its rightmost arm and connecting ribs (in what follows, we refer to
the remaining arm of the mechanical resonator as the 'optical arm').
We first generate Bloch bands for photonic crystals with unit cells
as highlighted in Fig.~\ref{FIG_B}(a), with optical arm widths
$W_{A}$ constant in $z$. As seen in Fig.~\ref{FIG_B}(b), the
fundamental TE Bloch band displays bandgaps at the $X$-point (bands
for $W_A/W_O = 0.25$ and $W_A/W_O = 0.65$ are displayed), and no
overlapping bands below the light-line in the spectral region near
the lowest band-edge. Figure~\ref{FIG_B}(c) shows the squared
electric field amplitude for the Bloch modes at the Brillouin zone
boundary, for the structure with $W_A = 0.65\cdot W_{O}$

Figure~\ref{FIG_B}(d) shows the evolution of the first four TE bands
at the $X$-point as a function of $W_A$. The apparent reduction in
bandgap (i.e., between the first and second bands) width with
increasing $W_A$ is due to an effective decrease in the filling
fraction, or the ratio between air and dielectric regions in the
unit cell. The larger bandgaps achieved with smaller $W_A$
correspond to a large effective reflectivity and thus are better
suited for strong spatial confinement. On the other hand, a wider
$W_A$ is desirable for enhanced modal frequency shifts with gap
width (i.e., optomechanical coupling). This is evident in
Fig.~\ref{FIG_B}(e), where $\lom$ is plotted, assuming
$\left|\mathbf{Q}\cdot\mathbf{n}\right|=1$ in eq.~(\ref{eq:Lom}),
for the fundamental TE photonic crystal modes at the $\Gamma$-point,
as a function of the $W_A/W_{O}$ ratio. A minimized optomechanical
length can be obtained for $W_A/W_{O}\approx 0.7$, about $30~\%$
lower than the maximum value within the plotted range. A trade-off
between spatial confinement and optomechanical coupling must thus be
achieved in a geometry where the mechanical beam width $W_A$ is
gradually reduced away from the cavity center, until a 'nominal'
width is reached, for which the reflectivity is highest. We thus
choose to let $W_A$ vary from $0.65\cdot W_O$ at the cavity center
to $0.25\cdot W_O$ at the mirrors.

The photonic bands shown in Fig.~\ref{FIG_F}(a) were obtained with
$w_z=0.35\cdot W_O$, $w_x=0.71\cdot W_O$, and $t=0.75\cdot W_O$ (
$w_z$ and $w_x$ are the elliptical holes' axes in the $z$ and $x$
directions, $t$ is the $\SiN$ thickness), parameters that led to a
large bandgap for $W_A=0.25\cdot W_O$. We choose the $\SiN$
thickness to be $t=350$~nm and the gap to be $w_\text{gap}=25$~nm,
and determine the remaining dimensions by regarding the cavity
center, where $W_O=0.65\cdot W_A$. A zero-finding routine was used
to determine the width $W_O$ and the corresponding cavity lattice
constant $a$ for the lowest $X$-point TE Bloch mode to be at a
wavelength $\lambda\approx980$~nm. We point out that while such
small gap width poses a great fabrication challenge, it may in
principle be achieved as in~\cite{ref:camacho}, where the $\approx$
GPa Si$_{3}$N$_{4}$ film stress was harnessed to bring two nanobeams
together as closely as 40~nm.

\begin{figure}[]
\centerline{\includegraphics[width=13.5cm,trim=65mm 0mm 50mm
0mm]{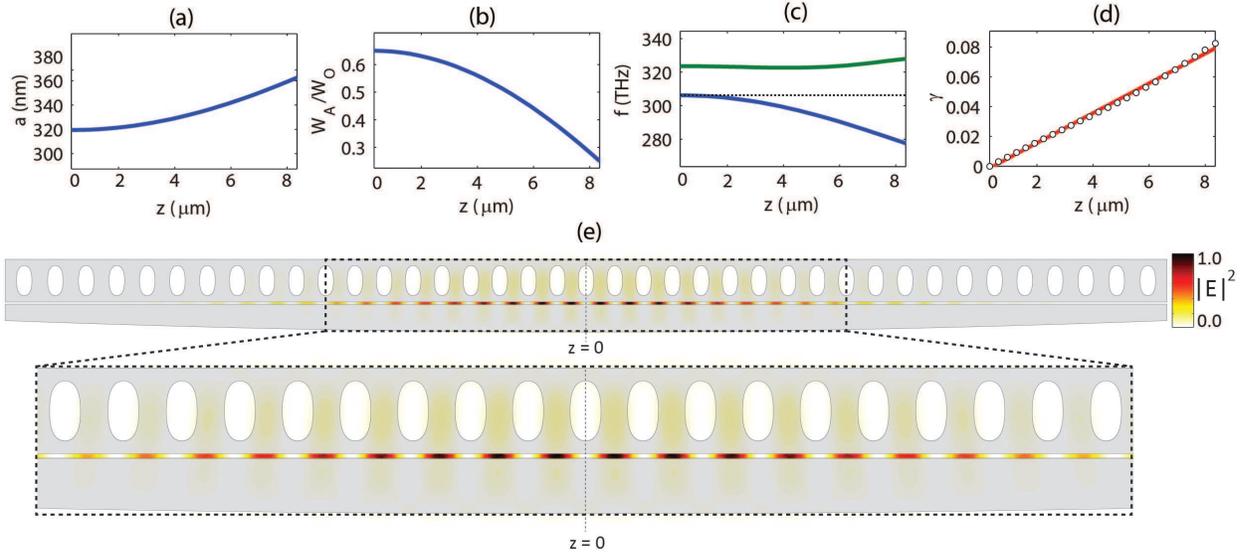}}\caption{ (a) Photonic lattice constant, (b)
optical arm ($W_A$) over optical beam ($W_O$) width ratio, (c)
frequency (at the X-point) of the first two TE photonic bands, and
(d) photonic mirror strength, as functions of the distance from the
cavity center ($z=0$). (e) Squared electric field amplitude for the
photonic mode generated with the parameters in (a)-(d).}
\label{FIG_C}
\end{figure}

To produce an effective potential well for the fundamental TE mode
at the $X$-point, the band edge must shift towards lower frequencies
at increasing distances from the cavity center. This shift is
opposite to what is obtained when $W_A$ is modulated in the desired
way, and thus we choose to modulate the lattice constant along the
cavity in order to produce the correct trend. Following the
procedure described in~\cite{ref:Quan1}, allowing both beam width
$W_A$ and the lattice constant $a$ to vary quadratically away from
the cavity center (Figs.~\ref{FIG_C}(a) and (b)), the desired trend
for the lowest band edge is achieved (Fig.~\ref{FIG_C}(c)), and an
approximately linear mirror strength (Fig.~\ref{FIG_C}(d)). The
mirror strength corresponds to the imaginary part of the Bloch
wavenumber for bandgap frequencies at the X-point,
$k=\pi/a(1+i\gamma)$. From 1D first order perturbation theory,
\begin{equation}
\gamma = \sqrt{\left(\frac{\omega_2-\omega_1}{\omega_1+\omega_2}\right)^2
-\left(\frac{\omega-\omega_0}{\omega_0}\right)^2},
\end{equation}
where $\omega_{1,2}$ are the dielectric and air band edges at the
Brillouin zone boundary, and $\omega_0$ is the midgap frequency.
Linear mirror strength profiles tend to produce optical modes with
reduced spatial harmonics above the light line. This leads to
reduced power leakage into the air, and thus higher quality
factors~\cite{ref:Quan1}. We point out that the quadratic lattice
constant and beam width tapers employed here, though not optimal,
produce sufficiently linear mirror strength profiles for high
quality factors to be achieved.

Figure~\ref{FIG_C}(e) shows a cavity mode at a wavelength
$\lambda=976$~nm, calculated with a full vector finite element
method, with mechanical beam width and lattice constant modulation
in Fig.~\ref{FIG_C}(a). The field is strongly concentrated in air
gap, as in Fig.~\ref{FIG_B}(c), and the quality factor is $Q =
12\times10^6$.

\subsection{Mechanical resonator design}
\label{section:mechanical} The optical resonator design dictates the
geometry of the mechanical beam's optical arm. The remaining parts
of the mechanical resonator (Fig.~\ref{FIG_F}) are designed around
this constraint. While the secondary arm can in principle be chosen
arbitrarily, for simplicity we let it be identical to the optical
arm. The lattice formed by the connecting ribs are made to follow
the optical lattice. This configuration minimized deterioration of
the optical quality factor due to scattering at the ribs. Rib widths
seemed to have a strong effect on scattering, and had to be made
small ($\leq50$~nm) to produce sufficiently low $Q$ degradation.

The mechanical resonator geometry studied here is similar to the
optomechanical nanobeam crystal analyzed in~\cite{ref:eichenfield3}.
While in this reference a detailed account of the phononic crystal
mechanical band structure is provided, here we focus only on the
relevant mechanical bands for enhanced optomechanical coupling. We
are interested in producing 'breathing' mechanical modes which
displace the nanobeam arms laterally, allowing the air gap width to
change.

\begin{figure}[b]
\centerline{\includegraphics[width=13.5cm]{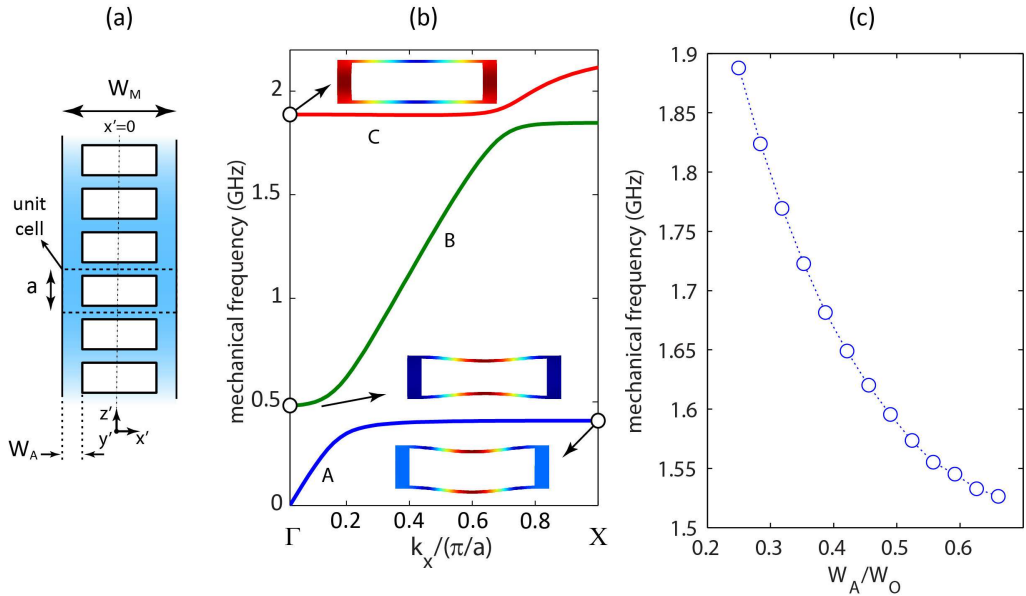}}\caption{(a)
Phononic crystal geometry. (b) Mechanical Bloch mode band structure
for a phononic crystal with $W_A/W_O=0.25$. The inset shapes show a
top view of the unit cell, with exaggerated boundary displacements
for modes on bands A, B and C. (c) Evolution of the $\Gamma$-point
eigenfrequency for band C, as a function of the ratio between the optical
arm ($W_A$)  and optical beam ($W_O$) widths.} \label{FIG_G}
\end{figure}

Phononic bands for the crystal shown in Fig.~\ref{FIG_G}(a) are
plotted in Fig.~\ref{FIG_G}(b), for an arm width $W_{A}=0.25\cdot
W_O$. Bands A, B and C were obtained with symmetric boundary
conditions for the displacement at the $x'=0$, $y'=0$ and $z'=0$
planes, so that no modes with other symmetries are displayed. We
refer to ref.~\cite{ref:eichenfield3} for a more detailed account of
the additional existing bands. The characteristic boundary
displacements for each band are inset in Fig.~\ref{FIG_G}(b). Near
the $\Gamma$-point, the mechanical modes of band C display the
desired lateral displacement pattern, necessary for the formation of
breathing resonances.

\begin{figure}[t!]
\centerline{\includegraphics[width=13.5cm,trim=60mm 0mm 60mm
15mm]{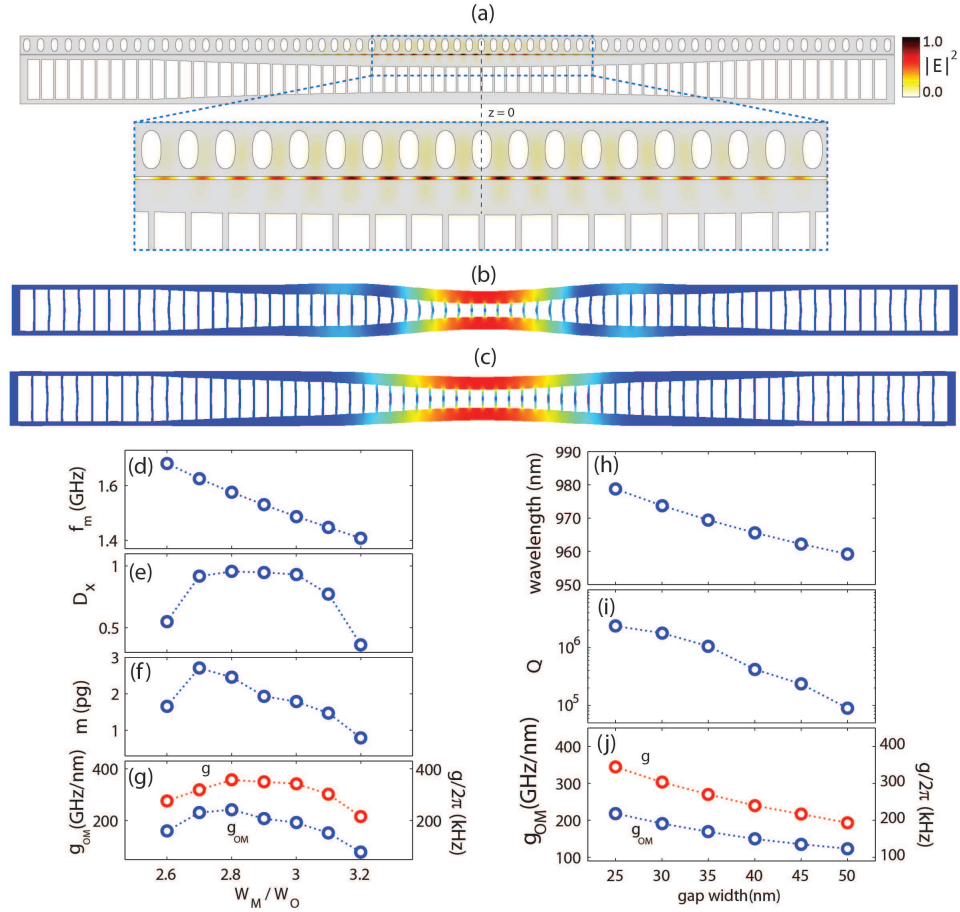}}\caption{ (a)Fundamental optical gap
resonance. (b) and (c): Mechanical beam breathing modes for(b)
$W_M/W_O=2.6$ and (c) $W_M/W_O=2.8$. (d) Mechanical frequency, (e)
normalized displacement $D_x$, (f) motional mass $m$ and (g) $g$ and
$\gom$ for the fundamental breathing mode as a function of
$W_M/W_O$. As a function of gap width: (h) fundamental optical
resonance wavelength, (i) quality factor and (j) optomechanical
couping ($\gom$) and coupling rates ($g$) with the fundamental
mechanical breathing mode for $W_M/W_O=2.8$ of (b). Note that the
geometry that produced (h)-(j) was optimized for a 25~nm gap.
Improved values of $\gom$, $g$ and $Q$ than plotted can in be
obtained with proper optimization for different gap widths (see
text).} \label{FIG_OMC}
\end{figure}

Modes from bands A and B stem from flexural modes of the connecting
ribs - evident in the inset displacement plots- and display
negligible lateral displacement. Band C has a minimum at $k_x=0$,
which creates the conditions for the formation of a breathing mode
phononic resonance. Indeed, a phononic cavity naturally arises from
the quadratic spatial arm width ($W_{A}$) variation, as evidenced in
Fig.~\ref{FIG_G}(c), where the $\Gamma$-point edge of band C is
plotted as a function of $W_{A}/W_O$. As the arm width decreases
from the cavity center towards the mirror regions, the band edge
moves towards higher frequencies, placing the resonance frequency
within the breathing mode phononic bandgap. Localized breathing
modes as plotted in Figs.~\ref{FIG_OMC}(a) and (b) result, with
frequencies that decrease with increasing nanobeam widths $W_M$, as
shown in Fig.~\ref{FIG_OMC}(c). It is also apparent that
displacement profiles can also change considerably with $W_M$. This
is quantified in Fig.~\ref{FIG_OMC}(d), where the normalized
displacement
\begin{equation}
D_x = \frac{\int{dA\mathbf{Q}\cdot \mathbf{n}}}{\int{dA{\left|\mathbf{Q}\cdot
\mathbf{n}\right|}}},
\end{equation}
is plotted as a function of $W_M/W_O$. The integral is performed
over the optical arm's surface facing the optical beam, and so $D_x$
is a measure of the uniformity of the optical arm's displacement
towards the optical beam. The evolution of the modal displacement
pattern also results in a variation of the motional mass, as shown
in Fig.~\ref{FIG_OMC}(e)

\subsection*{Optomechanical coupling}
The optomechanical coupling $\gom$ and coupling rate $g$ between the
optical resonance calculated in Section~\ref{section:optical} and
the breathing mode mechanical resonances discussed in
Section~\ref{section:mechanical} are plotted in
Fig.~\ref{FIG_OMC}(f), as a function of $W_A/W_M$. The maximum
$\gom$ and $g$ values achieved for the $W_M/W_O$ range shown are
$>2\times$ the value obtained with the optimized single nanobeam of
Section~\ref{section:single_nanobeam}, despite the larger motional
masses (Fig.~\ref{FIG_OMC}(e)). The strong influence of the gap
width on the optomechanical coupling is evident in the fact that
both $g$ and $\gom$ approximately follow the trend of $D_x$. In
fact, the effect of increasing gap widths on the resonance
wavelength is plotted in Fig.~\ref{FIG_OMC}(g), displaying a fast
blue shift for decreasing gaps. Because the cavity was initially
designed to have a 25~nm gap, the quality factor decreases as this
parameter is changed, and so do the optomechanical coupling $\gom$
and the coupling rate $g$, despite the increase in the optical
frequency $f_o$ ($g\propto x_{zpf}\propto f_o$,
$x_{zpf}=\sqrt{\hbar/2m\omega_m}$ is the zero point fluctuation
amplitude for mechanical displacement). For any fixed gap width,
however, the hole lattice can be adjusted to yield a high quality
factor. For instance, changing the the lattice constant profile in
Fig.~\ref{FIG_C}(a) so it varies (quadratically) from 325~nm at
$x=0$ to 365~nm at the cavity edge gives an optical mode at
wavelength $\lambda=986.4$~nm with $Q=2.4\times 10 ^{6}$ and
$\gom=180$~GHz/nm. In contrast, for the 25~nm gap design of
Fig.~\ref{FIG_OMC}, a 40~nm gap yields $\lambda=965.6$~nm,
$Q<5\times10^5$ and $\gom=150$~GHz/nm. This serves to show that our
double nanobeam design affords great flexibility towards optimized
performance under constraints such as resonance wavelength, quality
factor or gap width.

\subsection{Effect of index contrast}
Equation~\ref{eq:Lom} suggests that the higher index contrasts
generally yield higher optomechanical coupling. In addition, the
electric field concentration in the slot region ($\propto (\Delta
n)^2$~\cite{Almeida.OL.04}) is expected to add a significant
contribution to $\gom$ and $g$. Indeed, a design for operation at
the 1550~nm band based on silicon ($n_\text{Si}\approx3.48$),
produced with the same procedure as above, yielded an optomechanical
coupling rate~$g/2\pi\approx900$~kHz for a mechanical resonance at
$f_m=1.38$~GHz (an optimized single nanobeam design with similar
geometrical parameters displayed
 $g/2\pi\approx 400$~kHz for a mechanical resonance at $f_m=4.3$~GHz).
Table~\ref{table:OMC_for_Si_nanobeam} shows the relevant
optomechanical quantities for this design. Here, the 1D lattice
constant was allowed to vary quadratically from 360~nm to 385~nm,
and the optical beam width varied between from 325~nm at the cavity
center to 150~nm.

\begin{table}
\caption{Silicon Based Optical Resonator Optomechanical Crystal
Parameters.  A slot width of 10~nm was assumed.} \centerline{
    \begin{tabular}{|c|c|c|c|c|c|c|}
        \hline
        $\lambda_o$ (nm) &$f_m$(GHz) & $Q\times10^6$ & $\gom$ (GHz/nm) & $g/2\pi$ (kHz) & $\lom$ ($\mum$) & m (pg) \\ \hline
        \hline
        1540.4 & 1.38 & $1.5$ & 485 & 882 & 0.4 & 1.8 \\ \hline
    \end{tabular}
}    \label{table:OMC_for_Si_nanobeam}
\end{table}

\section{Applications}
\label{section:Applications}

\subsection{Wide spectral separation wavelength conversion  }
\label{section:wlc_double} While a 'standard' nanobeam geometry such
as that studied in Section~\ref{section:single_nanobeam} already
offers the necessary conditions for wavelength conversion, the
spectral separation between optical resonances of different orders
is limited to, at best, 70~nm. Even in silicon nanobeam designs,
where the higher index contrast leads to a considerably wider
spacing between resonances, a maximum separation of $\approx 100$~nm
is achievable~\cite{ref:hill_WLC}. As discussed in
Section~\ref{section:single_nanobeam}, the achievable wavelength
separation in the single nanobeam design is ultimately given by the
nanobeam photonic bandgap, which can be limited even for high
contrasts such as in the Si case. The ability to perform wavelength
conversion over considerably wider wavelength separations may be
desirable for applications in which classical or quantum information
transmitted over optical fibers at telecom wavelengths is originally
produced at completely separate wavelength bands, for instance in
the near-infrared or visible range. Such an achievement would be
enabling for proposed hybrid quantum optical networks, whose nodes
may be widely different from each other (trapped
atoms~\cite{ref:Kimble_Nat08}, quantum
dots\cite{ref:Rakher_NPhot_2010}, organic
molecules~\cite{ref:Hwang_J}, etc).

\begin{figure}[]
\centerline{\includegraphics[width=13.5cm,trim=0mm 20mm 0mm
0mm]{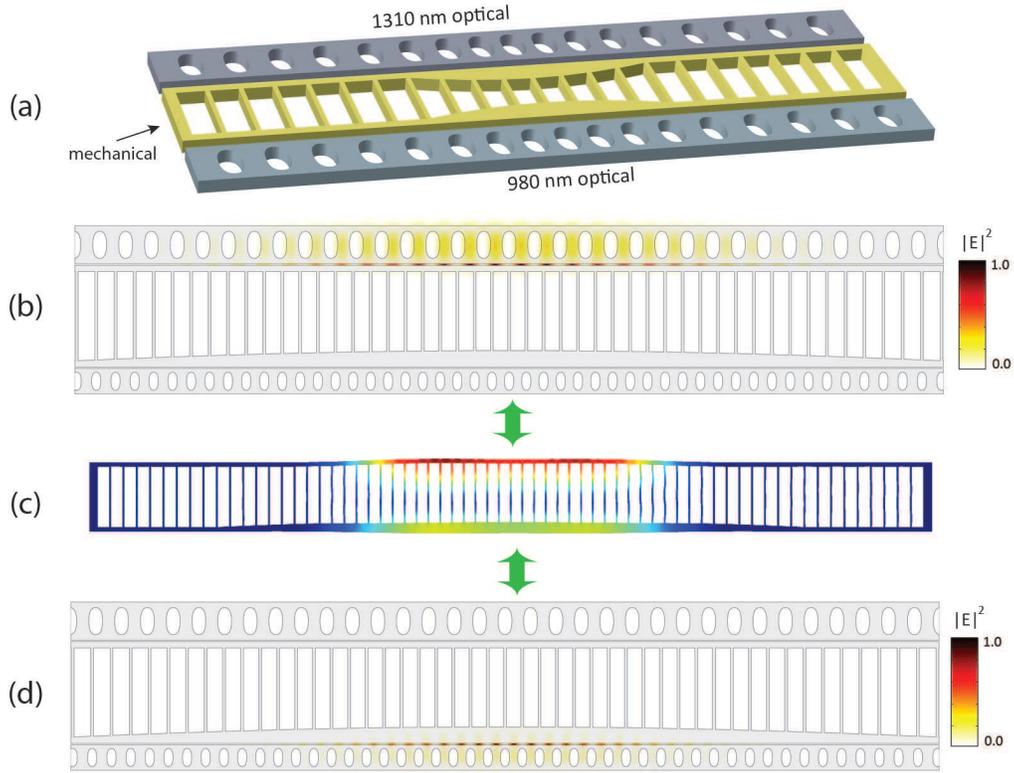}}\caption{(a) Double optical cavity
optomechanical crystal geometry. (b) 1310~nm optical mode. (c)
Mechanical resonator breathing mode. (c) 980~nm optical mode. Green
arrows indicate simultaneous  coupling between the optical
resonances and the mechanical mode.} \label{FIG:double_beam}
\end{figure}

The physical separation between optical and mechanical resonances
afforded through our double nanobeam optomechanical approach allows
the formation of optomechanical resonators with widely spaced,
spatially separate and semi-independent optical resonances with
reasonable optomechanical coupling to the same mechanical resonance.
Such a resonator, schematically depicted in
Fig.~\ref{FIG:double_beam}(a), consists of a central mechanical
nanobeam laterally sandwiched by two optical nanobeams. Air gaps on
the two sides of the mechanical resonator allow a single breathing
mode resonance to couple efficiently to optical nanoslot modes on
both sides. We demonstrate the feasibility of this scheme via an
example in which optical resonances at wavelengths of 980~nm and
1310~nm are supported. Figures~\ref{FIG:double_beam}(b) and (d)
respectively show the 1310~nm and 980~nm resonances of a double
nanobeam resonator that share the mechanical resonance at
$f_m=1.38$~GHz in Fig.~\ref{FIG:double_beam}(c). Both optical
resonances display quality factors in excess of $10^6$, which allows
operation in the resolved sideband regime. The 980~nm wavelength
optical cavity has exactly the same geometry parameters as that in
Section~\ref{section:optical}. The 1310~nm cavity was designed with
the same procedure as described in Section~\ref{section:optical}.
Achieving high optical quality factors for both cavities required
significant modifications to the original mechanical nanobeam
geometry presented in Section~\ref{section:mechanical}. First, the
nanobeam width $W_M$ had to be increased to $4\cdot W_O$, to
minimize the interaction between the two optical resonators. This
resulted in a lower resonance frequency, $f_m=1.38$~GHz. In
addition, the 1310~nm resonator's optical arm was made constant with
$W_A=0.25\cdot W_O$, to prevent coupling of the confined 980~nm
resonance to leaky resonances of the 1310~nm beam. This caused the
optomechanical coupling $\gom$ for the 1310~nm resonator to be
$\approx40~\%$ lower than that obtained with modulation of the
optical arm ($\gom\approx112$~GHz/nm). It is worth noting on the
other hand that, despite the uniform arm width, $\gom$ can still be
reasonably high. This parameter, together with the optomechanical
coupling rate $g$, effective optomechanical length $\lom$ and
motional mass $m$ for the two resonators are given in
Table~\ref{table:OMC_for_double_nanobeam}.

\begin{table}
\caption{Double optical Resonator Optomechanical Crystal Parameters.
Both slots are assumed to be 25 nm wide. } \centerline{
    \begin{tabular}{|c|c|c|c|c|c|}
        \hline
        $\lambda_o$ (nm) & $Q\times10^6$ & $\gom$ (GHz/nm) & $g/2\pi$ (kHz) & $\lom$ ($\mum$) & m (pg)  \\ \hline
        \hline
        978 & $4.4$ & 179 & 322 & 1.7 & 1.92  \\ \hline
        1318 & $4.6$ & 67 & 122 & 3.4 & 1.85  \\
        \hline
    \end{tabular}
}    \label{table:OMC_for_double_nanobeam}
\end{table}

As discussed in the Appendix, the conversion efficiency achievable
in the scheme proposed in ref.~\cite{ref:safavi-naeini3} and
demonstrated in ref.~\cite{ref:hill_WLC} depends on the
cooperativities $C_{1,2}\propto g_{1,2}^2n_{c;1,2}$ of optical
cavities 1 and 2 ($n_{c;1,2}$ is the average photon population in
each cavity), and unity conversion requires $C_1=C_2\gg1$.  The
coupling rates calculated above are approximately a factor of three
times smaller than those used in ref.~\cite{ref:hill_WLC}. These
smaller coupling rates can be compensated by increasing the number
of photons in the cavity (by a factor of nine, assuming similar
optical and mechanical decay rates are achievable), which should be
feasible in the Si$_3$N$_4$ system.  We also note that the disparate
coupling rates for the two cavities is not expected to influence the
ability to achieve high efficiencies, as such differences may be
compensated by proper balancing of the intra-cavity photon
populations $n_{c;1,2}$.

\subsection{Strong optomechanical coupling}
The observation of quantum behavior in cavity optomechanics relies
on the possibility of strong coupling between optical and mechanical
resonances. Ultimately, a regime is sought in which interactions at
the single photon and phonon levels are observable. This
single-photon strong coupling regime, in which a single phonon is
able to shift an optical resonance by an extent comparable to the
latter's linewidth, is characterized by a coupling rate $g$
comparable to both the optical resonance decay rate $\kappa$ and the
mechanical frequency $\Omega$. Achieving single photon strong
coupling at optical frequencies is challenging due to high optical
losses. In~\cite{ref:Ludwig}, a scheme was proposed for producing a
strongly enhanced, effective optomechanical coupling for quantum
non-demolition (QND) photon detection and phonon number readout
measurements. Such a QND measurement would extract information from
the quantum system without disturbing its quantum state. In the
setup of~\cite{ref:Ludwig}, an optomechanical crystal geometry
supports two optical resonances split by a frequency $2J$,
comparable to the mechanical frequency $\Omega$ of an interacting
mechanical resonance. The coherent interaction between photons and
phonons is characterized by an effective coupling rate
$g_0^2/\delta\Omega$, where $\delta\Omega=2J-\Omega$ and $g_0$ is
the bare optomechanical coupling rate. In the limit
$2J\rightarrow\Omega$, the optical frequency shift is
$g_0^2/\delta\Omega n_b$, where $n_b$ is the phonon number,which may
be sufficiently large to produce a phonon number readout.

\begin{figure}[]
\centerline{\includegraphics[width=13.5cm]{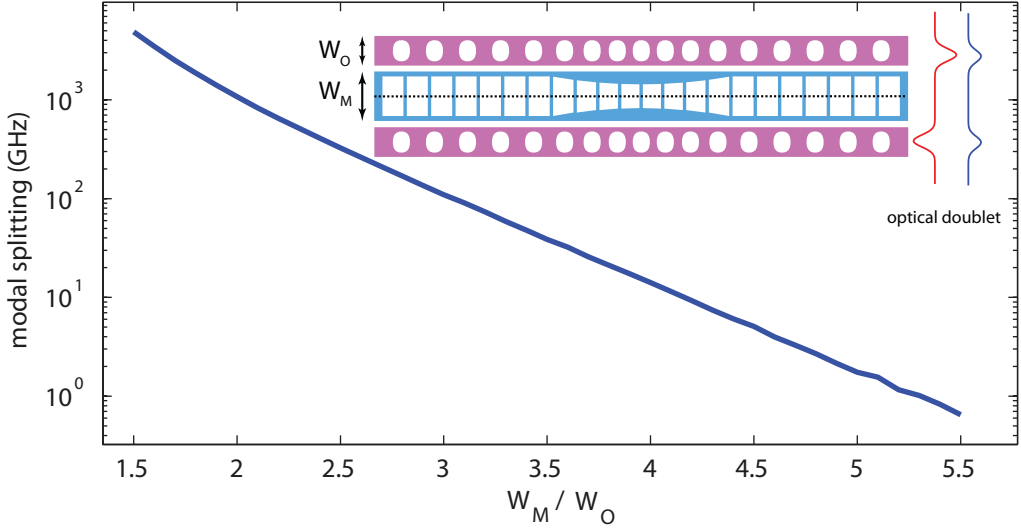}}
\caption{Optical mode splitting for the symmetric double optical beam resonator in the
inset. The dotted line in the inset indicates the symmetry plane that defines
splits the original optical resonance into symmetric (blue) and anti-symmetric (red)
modes.
Parameters for the optical and mechanical resonators are as
in Section~\ref{section:gap_nanobeams}, with $W_M/W_O=2.8$}.
\label{FIG:double_nanobeam_strong_coupling}
\end{figure}

The flexibility of our double nanobeam geometry allows us to
accommodate the optical mode splittings $2J\approx\Omega$ required
in the scheme proposed in~\cite{ref:Ludwig}. As shown in the inset
of Fig.~\ref{FIG:double_nanobeam_strong_coupling}, an optomechanical
resonator is formed by two identical optical beams sandwiching a
mechanical resonator. The two individual optical resonances couple
evanescently, forming a symmetric / anti-symmetric doublet whose
spectral separation can be controlled by the mechanical beam width
$W_M$. The modal splitting $2J$, shown in
Fig.~\ref{FIG:double_nanobeam_strong_coupling} for the same cavity
parameters as in Section~\ref{section:single_nanobeam}, indeed
decreases with increasing $W_M$, as the spatial overlap between the
individual optical resonances diminishes. The main breathing mode
mechanical frequency also decreases with $W_M$, however remaining
within the GHz range even for the largest mechanical beam widths
plotted, so that the regime $\delta\Omega=2J-\Omega\to0$ can be
achieved. The coupling between the two modes can also be tuned by
creating optical cavities with coinciding individual resonances, but
different gap widths.

\subsection{Microwave to optical photon conversion}
State-of-the art superconducting microwave circuits can be designed
to provide controllable, coherent interactions between
superconducting qubits and high quality microwave cavity resonators,
and provide favorable conditions for the creation and manipulation
of photonic states at microwave frequencies. In the context of a
quantum network~\cite{ref:Kimble_Nat08} where the nodes are
connected via photonic links, however, low loss transmission of
microwave photons over long distances becomes a challenge, as
superconducting transmission lines would be required. The ability to
convert between microwave and optical photons which can be
transmitted via optical fibers thus constitutes an appealing
solution to this problem.

\begin{figure}[]
\centerline{\includegraphics[width=13.5cm]{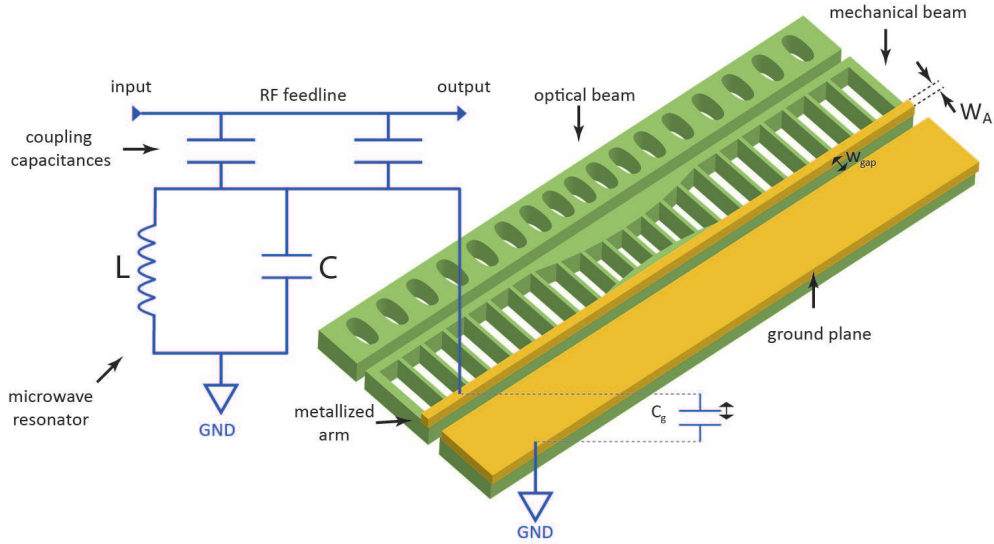}}
\caption{Schematic for possible optomechanical microwave to optical
wavelength converter. The RF cavity resonance frequency is modulated
by the gap capacitance $C_\text{g}$, which depends on the gap width,
and therefore on the displacement of the mechanical resonator. The
latter is coupled to the slot optical mode formed with the optical
beam.} \label{FIG:optical_to_RF}
\end{figure}

The physical separation between optical and mechanical resonances in
our geometry creates opportunities for the generation of
microwave-to-optical frequency signal transducers. One possible
geometry is illustrated in Fig.~\ref{FIG:optical_to_RF}, where the
far arm of the mechanical resonator incorporates a metallic strip,
and, together with a neighboring ground plane, forms a capacitor
that integrates a superconducting, planar circuit microwave
resonator, similar to that demonstrated
in~\cite{ref:regal,ref:Sulkko}. The displacement of the mechanical
beam affects the cavity resonance by causing a variation in the
circuit capacitance. Coupling between the mechanical displacement
and cavity capacitance is stronger for smaller gaps between the
metallized beam and the ground plane. At the same time, the distance
between the metallized mechanical beam arm and the optical resonator
must be sufficient to prevent significant deterioration of the
quality factor of the optical cavity, while keeping the
superconducting microwave resonator sufficiently separated from
strong optical fields is necessary to avoid heating and degradation
of its performance. As shown in
Section~\ref{section:single_nanobeam}, this separation can be
accomplished through increased mechanical beam widths.  While this
leads to decreasing mechanical frequencies for the fundamental
breathing mode, as seen in the previous sections, mechanical
frequencies in the GHz range are achievable.

As is the case with optical-to-optical conversion,
microwave-to-optical conversion under the protocol described in
ref.~\cite{ref:safavi-naeini3} requires
$C_{\mu\text{wave}}=C_{\text{optical}}\gg1$, that is, the
cooperativity of the microwave and optical resonators (see Appendix)
should be large and matched to each other. We anticipate that the
design of the optical resonator will be similar to that described
previously, while a full design of the microwave cavity is beyond
the scope of this paper. Qualitatively, we note the conceptual
similarity between the geometry shown in
Fig.~\ref{FIG:optical_to_RF} and that studied experimentally in
ref.~\cite{ref:Sulkko}, where a suspended Al nanobeam separated from
a ground plane by a gap $\approx20$~nm was employed to perform
vibration detection near the quantum limit, and for which subsequent
experiments have shown radiation-pressure driven phenomena like
electromagnetically-induced transparency~\cite{ref:Massel}. The
coupling between the microwave cavity and mechanical resonator is
$g_{\mu\text{wave}}\sim(\partial C_\text{g}/\partial
w_{\text{gap}})$, where $C_{\text{g}}$ is the coupling capacitance
and $w_{\text{gap}}$ is the gap between the two elements. Reaching
the regime $C_{\mu\text{wave}}\gg1$ with a microwave intracavity
photon number consistent with recent experiments (up to 10$^6$
photons were used in ref.~\cite{ref:Massel}) will thus require small
gaps (10~nm gaps were reported in ref.~\cite{ref:Sulkko}), along
with low dissipation for both the microwave cavity and the
mechanical resonator. Investigation of such suitable microwave
cavity geometries is in progress.

\section{Conclusion}
We have presented a design methodology and analysis of an
optomechanical crystal comprised of two laterally coupled 1D
nanobeams which support spatially separate optical and mechanical
resonances. This spatial separation enables independent design and
flexible optimization of the optics and mechanics of the system, and
can be of particular importance in applications requiring the
coupling of multiple electromagnetic modes to a single localized
mechanical resonance. In addition, the small gap ($\approx$25~nm)
separating the two nanobeams gives rise to a slot optical mode
effect that enables a large zero-point optomechanical coupling
strength to be achieved, even for nanobeam materials with small
refractive index contrasts such as Si$_3$N$_4$. Our design predicts
optical quality factors above $10^6$ and mechanical frequencies
above $1$~GHz for nanobeams in both the Si$_3$N$_4$ and Si systems.
While the optomechanical coupling strengths predicted here are
moderately superior to those calculated and experimentally
determined from single nanobeam geometries, optical and mechanical
resonances can be designed with considerably more independence and
control.

The large predicted optomechanical coupling strengths to GHz
mechanical oscillators in Si$_3$N$_4$ are particularly attractive,
as this material displays a broad optical transparency window, which
allows operation throughout the visible and near-infrared. The
material also offers a low intrinsic mechanical dissipation rate,
and does not exhibit the two-photon absorption and subsequent
free-carrier absorption and dispersion observed in
silicon~\cite{ref:Barclay7}, potentially allowing higher powers to
be employed and increasing the range of achievable pump-enhanced
optomechanical coupling values. All of these features provide good
prospects for the realization of radiation-pressure mediated
photon-phonon translation ~\cite{ref:safavi-naeini3}.

\section*{Acknowledgements}
This work has been supported in part by the DARPA MESO program. We
thank Vladimir Aksyuk and Jeff Hill for valuable discussions.

\renewcommand{\theequation}{A-\arabic{equation}} % redefine the command that creates the equation no.
\setcounter{equation}{0}  % reset counter
\section*{Appendix - Optomechanical wavelength conversion}
The wavelength conversion scheme proposed in
Sections~\ref{section:wlc_single} and~\ref{section:wlc_double}
involve two optical resonances at frequencies $\omega_{o,1}$ and
$\omega_{o,2}$, both of which are coupled to a mechanical resonance
at frequency $\omega_m$. Both optical cavities are prepared with
strong pumps, red-detuned from the resonance centers by the
mechanical mode frequency. For each optical resonance, this
preparation gives rise to an effective optomechanical interaction
{\it for signals at the cavity center} described by a beam-splitter
type interaction Hamiltonian $H_\text{int,eff}=\hbar
G\left(\hat{a}^\dagger\hat{b} + \hat{a}\hat{b}^\dagger\right)$,
where $\hat{a}$ and $\hat{b}$ are the destruction operators for
cavity photons at the center frequency and phonons in the mechanical
resonator. The parameter $G$ is the pump-enhanced optomechanical
coupling, $G=g|\alpha^{ss}|$, where $\alpha^{ss}$ is the square root
of the steady-state photon population at the pump frequency, and $g$
the bare-cavity optomechanical coupling rate. It is apparent from
the effective Hamiltonian that the interaction allows the exchange
of signal photons and phonons with a rate dictated by the intensity
of the pump beam. In the wavelength conversion process, a signal
photon at frequency $\omega_1$ is injected into the (prepared)
cavity 1, and converted to a cavity phonon. The process is
reversible, such that the newly created phonon may be subsequently
transduced to a photon in (also prepared) cavity 2, at frequency
$\omega_{o,2}$. This process can be described classically via the
matrix equation~\cite{ref:hill_WLC}
\begin{equation}
\left[\begin{array}{c} \alpha_1 \\ \beta \\ \alpha_2
\end{array}\right] =
\left[
\begin{array}{c c c} -\kappa_1/2 & -iG_1 & 0 \\-iG_1 & -i\gamma_i/2 &
-iG_2 \\ 0 & -iG_2 & \kappa_2/2
\end{array} \right]^{-1}
\left[\begin{array}{c} \sqrt{\frac{\kappa_{ex,1}}{2}}\alpha_\text{in;1} \\
0 \\ 0 \end{array}\right], \label{eq:ppt}
\end{equation}
obtained from the Heisenberg equations for the bosonic operators
$\hat{a}_{1,2}$ (for photons in cavities 1 and 2) and $\hat{b}$. To
arrive at eq.~(\ref{eq:ppt}), we replaced
$\hat{a}_{1,2}\rightarrow\alpha_{1,2}\text{e}^{-i(\omega_{c;1,2}-\Delta
t)}$, $\hat{\beta}\rightarrow\beta\text{e}^{-i\Delta t}$, where
$\Delta$ is the detuning of the optical signal from the cavity
center frequency $\omega_{c;1,2}$, and the rotating wave
approximation was used. Noise sources were completely disregarded.
In eq~(\ref{eq:ppt}), $\kappa_{1,2} =
\kappa_{ex;1,2}+\kappa_{i;1,2}$ are the decay rates for optical
modes 1 and 2, comprised of intrinsic ($\kappa_i$, associated with
e.g., radiative losses), and extrinsic ($\kappa_{ex}$, associated
with coupling to external access channels) components; $\gamma_i$ is
the intrinsic decay rate for the mechanical mode; $\alpha_\text{in}$
and is the input optical field. Finally, it was assumed that the
detuning between the pump and signal beams (at the input and output
optical cavities) was equal to the mechanical frequency,
$\Delta=\omega_m$. The field output from cavity 2 is simply
$\alpha_\text{out;2}=\sqrt{\kappa_{ex,2}/2}\alpha_2$, so
eq.(\ref{eq:ppt}) can be solved to give
\begin{equation}
\alpha_\text{out;2} =
\sqrt{\eta_1\eta_2}\frac{\sqrt{\gamma_\text{OM;1}\gamma_\text{OM;2}}}{\gamma/2}\alpha_\text{in,1},
\end{equation}
where $\gamma = \gamma_\text{OM;1}+\gamma_\text{OM;2}+\gamma_i$ and
$\gamma_\text{OM;1,2} = 4G_{1,2}^2/\kappa_{1,2}$
($\gamma_\text{OM;1,2}$ are the optical spring contributions to the
total mechanical mode damping $\gamma$), and
$\eta_{1,2}=\kappa_{ex;1,2}/2\kappa_{1,2}$.The conversion
efficiency, then, is
\begin{equation}
\eta = \left|\frac{\alpha_\text{out,2}}{\alpha_\text{in,1}}\right|^2
= \eta_{1}\eta_{2}\frac{4C_1C_2}{(1+C_1+C_2)^2},
\end{equation}
where $C_{1,2}=4G_{1,2}^2/\kappa_{1,2}\gamma_i$ are the
cooperativities for optical cavities 1 and 2. It is apparent that,
for $\eta\to1$, $C_1=C_2\gg1$ and $\eta_{1,2}\to1$ are sufficient. A
more complete derivation of the wavelength process,including
treatment of noise, is given in~\cite{ref:hill_WLC}.

\end{document}